\documentclass[accepted]{article}

\usepackage{microtype}
\usepackage{graphicx}
\usepackage{subfigure}
\usepackage{booktabs} %
\usepackage{amsmath}

\PassOptionsToPackage{hyphens}{url}\usepackage{hyperref}

\usepackage[accepted]{mlsys2025}

\mlsystitlerunning{Optimizing PyTorch Inference with LLM-Based Multi-Agent Systems}

\usepackage{eso-pic}
\usepackage{xparse}

\newlength{\badgewidth}
\newlength{\badgegap}
\newcommand{\badgeList}{}

\NewDocumentCommand{\addTopRightBadge}{O{} m}{%
\gappto{\badgeList}{\href{#1}{\includegraphics[width=\badgewidth]{#2}}\hspace{\badgegap}}%
}

\newcommand{\placeTopRightBadges}{%
\AddToShipoutPictureBG*{%
\put(\LenToUnit{\paperwidth - 1.5cm - \badgewidth},\LenToUnit{\paperheight - 2cm}){%
\makebox[0pt][r]{\badgeList}%
}%
}%
}

\addTopRightBadge{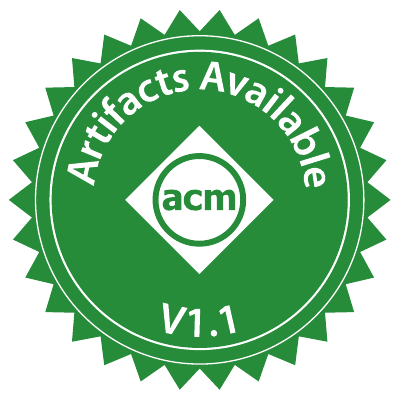}
\addTopRightBadge{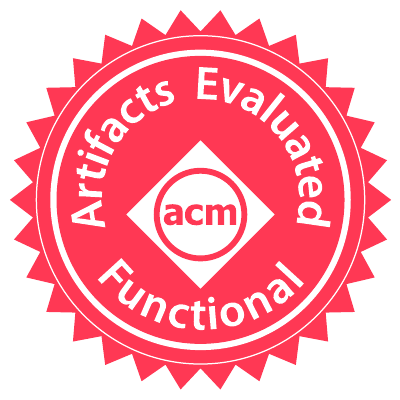}

\placeTopRightBadges

\begin{document}

\twocolumn[
\mlsystitle{Optimizing PyTorch Inference with\\LLM-Based Multi-Agent Systems}

\mlsyssetsymbol{equal}{*}

\begin{mlsysauthorlist}
\mlsysauthor{Kirill Nagaitsev}{nw}
\mlsysauthor{Luka Grbcic}{lbl}
\mlsysauthor{Samuel Williams}{lbl}
\mlsysauthor{Costin Iancu}{lbl}
\end{mlsysauthorlist}

\mlsysaffiliation{nw}{Northwestern University}
\mlsysaffiliation{lbl}{Lawrence Berkeley National Laboratory}

\mlsyscorrespondingauthor{Kirill Nagaitsev}{knagaitsev@u.northwestern.edu}

\mlsyskeywords{Machine Learning, AI, LLMs, Agents, PyTorch, GPUs, CUDA}

\vskip 0.3in

\begin{abstract}
Maximizing performance on available GPU hardware is an ongoing challenge for modern AI inference systems. Traditional approaches include writing custom GPU kernels and using specialized model compilers to tune high-level code for specific GPU targets. Recent work shows that LLM-based multi-agent systems can effectively perform such tuning, often outperforming existing compilers and eliminating the need for manual kernel development. However, the dynamics of multi-agent systems for this task remain unexplored. In this work, we present a logical framework for comparing multi-agent PyTorch optimization systems. Our evaluation shows that exploit-heavy strategies perform best when paired with error-fixing agents, and that performance correlates with the granularity of optimization steps. The best implementation achieves an average 2.88× speedup over PyTorch Eager (1.85× over torch.compile) on an H100 GPU across diverse tasks in KernelBench, a benchmark suite covering a range of machine learning architectures in PyTorch. Code is publicly available at: \url{https://github.com/pike-project/pike}

\end{abstract}
]

\printAffiliationsAndNotice{}  %

\section{Introduction}

One of the most crucial areas of performance optimization today is that of AI/ML workloads. As the landscape of GPUs rapidly evolves, AI/ML software must constantly adapt to capitalize on the performance benefits of new hardware. Unfortunately, the optimization of GPU kernels remains one of the most technically demanding frontiers in performance engineering, requiring mastery of GPU parallelism, memory hierarchies, and scheduling behavior.

Numerous model-level compilers enable fully automated AI/ML model optimization \cite{pytorch2, tensorrt, onnxruntime, xla}, but they require continual updates to target new GPUs and often fall short of hand-tuned performance. For instance, the original custom FlashAttention kernel \cite{dao2022flashattention} achieved a 7.6× speedup over a generic PyTorch implementation before being integrated as a built-in attention operator.

Domain-specific languages and libraries simplify manual GPU development and tuning \cite{kerr2017cutlass, tillet2019triton, spector2024thunderkittens}, yet achieving peak performance with tools like Triton \cite{tillet2019triton} still demands substantial time and expertise, including mastery of GPU memory hierarchies, tiling, and block sizing.

Recent work has applied LLM-based methods to the AI/ML GPU tuning problem. An influential effort is KernelBench \cite{ouyang2025kernelbench}, the first open benchmark for LLM‑generated GPU kernels, which frames the task as translating PyTorch operators into optimized kernels. Several studies have developed agent-driven systems for KernelBench \cite{lange2025ai, lange2025towards, li2025cuda, metr2025kernel, andrews2025gpu, baronio2025kevin} or related challenges, including direct kernel optimization \cite{novikov2025alphaevolve, wei2025astra}.

Despite this progress, the dynamics of multi-agent systems for this PyTorch/kernel search problem remain unexplored.
Specifically, the roles of individual agents, prompting strategies, and solution library designs have not been systematically examined.

LLM-based PyTorch/kernel optimization can be viewed as a search for an optimal solution, and an explore-exploit tradeoff emerges from this search process.
Exploration builds a broad base of valid solutions to avoid premature convergence, whereas exploitation concentrates effort on the best existing solutions.
We analyze how system components interact in multi-agent PyTorch optimization systems, focusing on the explore–exploit tradeoff, and identify an optimal configuration for PyTorch inference optimization.

To summarize, this work makes the following contributions:

\begin{itemize}
    \item We develop a logical framework for comparing multi-agent PyTorch optimization systems, along with our implementations within it, collectively defined as PyTorch Inference Kernel Evolution (PIKE).
    \item We develop PIKE-B, a multi-agent evolutionary branching strategy, along with PIKE-O, an OpenEvolve-based \cite{openevolve} strategy, finding that the exploit-heavy strategies with error fixing substantially outperform explore-heavy strategies.  
    \item We analyze the optimization trajectories and agent roles in these strategies, finding that performance correlates with the granularity of steps, with aggressive, exploit-heavy steps yielding better results within our budget.
    \item We achieve state-of-the-art speedups on a refined KernelBench suite \cite{ouyang2025kernelbench, metr2025kernel}, with best solutions implemented in both CUDA and Triton.

\end{itemize}

\section{Logical Framework}
\label{sec:logical-framework}

\begin{figure*}
    \centerline{\includegraphics[width=1\textwidth]{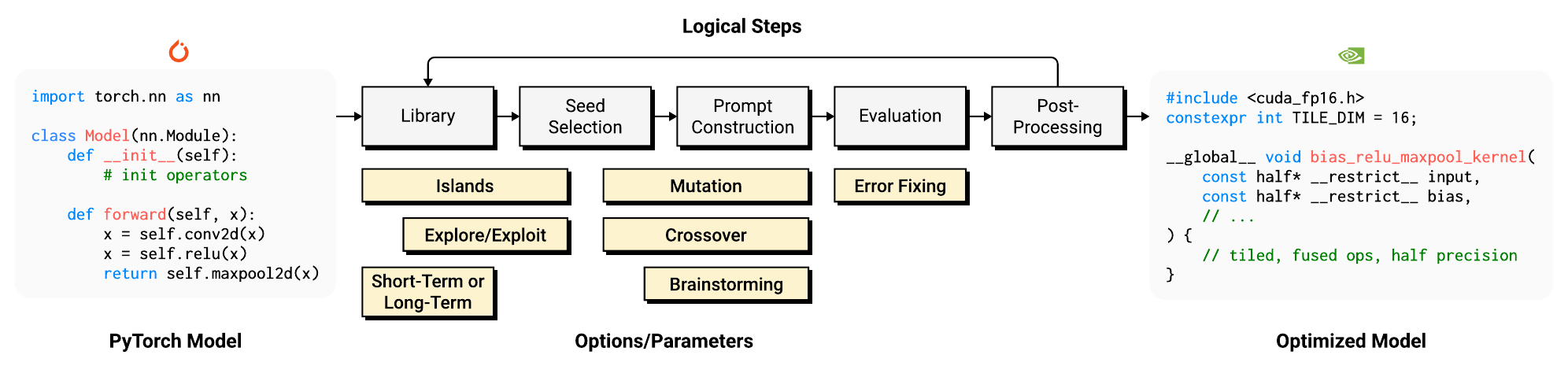}}
    \caption{Logical framework which multi-agent systems operate in, for the task of PyTorch inference optimization. Logical steps of the optimization process are shown above, and core options/parameters are shown below near the steps they relate to.}
    \label{fig:logical-framework}
\end{figure*}

Our objective is to translate PyTorch models into optimized versions that achieve inference speedups. To this end, we develop an LLM-based multi-agent logical framework and analyze its dynamics to identify the strategies that deliver the best performance.

Figure \ref{fig:logical-framework} shows the multi‑agent framework for optimizing PyTorch inference. It takes a PyTorch model as input and produces a translated, performance‑optimized version for efficient execution. Parameter choices within the framework either move the search process towards being exploitative, where the focus is on using the existing best model implementations found so far, or explorative, where a broad set of prior implementations are used to avoid premature convergence. The process is organized into several stages, each addressing a specific aspect of the optimization pipeline.

The process begins with the \textbf{library}, which stores the initial PyTorch model and valid solutions generated by the LLM, where a \textit{solution} is code generated by a single LLM invocation. The framework can loop multiple times, potentially adding new solutions to the library at each iteration. The library maintains either a \textit{long-term} or \textit{short-term} history of solutions, discarding the oldest entries after a set cutoff. Additionally, it may be organized into \textit{islands}, where each island has its own separate history of solutions. Islands are a concept from island-based genetic algorithms \cite{tanese1989islands}, and are used by numerous LLM-based agent frameworks \cite{novikov2025alphaevolve, openevolve}. In the \textbf{library} phase, the \textbf{Initial Brainstorming Agent (IBA)} can take the problem statement and input code as a seed to generate a list of $n$ optimization ideas for agents to pursue.

In the \textbf{seed selection} phase, an island is chosen (if multiple exist), helping to determine the specific library segment sampled in this stage. The selection process uses an \textit{explore/exploit} ratio to randomly pick a \textit{seed} solution for the next LLM prompt. With a ratio of $\varepsilon : (1 - \varepsilon)$, the probability of selecting from all history is $\varepsilon$, while elite solutions are chosen with probability $1 - \varepsilon$. The \textit{explore/exploit} ratio interacts with other parameters to influence the system's overall behavior, determining whether it leans toward exploration or exploitation. For instance, increasing the number of islands enhances exploration by utilizing distinct solution pools.

During the \textbf{prompt construction} phase, an LLM prompt is developed to optimize the original PyTorch model. This can involve \textit{mutation} (using one solution) or \textit{crossover} (drawing from multiple solutions) from the library. The \textbf{Code Optimization Agent (COA)} takes the problem statement, the original model, and any additional seeds to generate an optimized version.  Though mutation typically implies small changes in traditional evolutionary algorithms, the LLM is not limited to minor adjustments. Additionally, an optional \textit{brainstorming} element can be included, incorporating ideas from the IBA or previous LLM queries. The prompt is then used to query the LLM, generating a new solution.

The \textbf{evaluation} stage entails compiling and running the generated solution. This includes verifying correctness (ensuring functional equivalence with the original model) and measuring performance based on a defined metric (e.g., speedup). If compilation or correctness checks fail, an \textit{error fixing} loop may be initiated, where the \textbf{Error Fixing Agent (EFA)} makes an LLM query to address code issues before reevaluating the solution.

The final stage, \textbf{post-processing}, involves sending the validated new solution to the library alongside evaluation metrics. If an exit condition is met (e.g., exceeding a maximum number of iterations), the absolute best solution (according to the objective metric) is returned.

Importantly, our logical framework supports implementations that can invoke iterations of the logical loop in parallel. In such cases, the seed selection from the library phase draws from solutions generated only by completed iterations.

As we refine our multi-agent logical framework for optimizing PyTorch models, it is crucial to explore the various configurable parameters and their interactions to maximize performance. Key questions arise: How do different settings of the explore/exploit ratio impact the overall efficiency of the optimization process? In what ways does the number of islands influence both exploration and exploitation dynamics? Additionally, what effect does varying the elite archive size have on performance outcomes? 

We must also investigate whether the inclusion of the EFA contributes to improved results and under what circumstances it is most beneficial. Understanding these interactions will help us determine optimal configurations for specific tasks and ensure that our framework effectively balances the trade-off between exploration and exploitation. By systematically addressing these configuration questions, we aim to enhance the adaptability and efficacy of our LLM-based approach in translating and optimizing PyTorch models.

\section{Evolutionary Strategies for PyTorch Optimization}

\subsection{PIKE-B: Branching Search}

\begin{figure}
  \centerline{\includegraphics[width=.95\columnwidth]{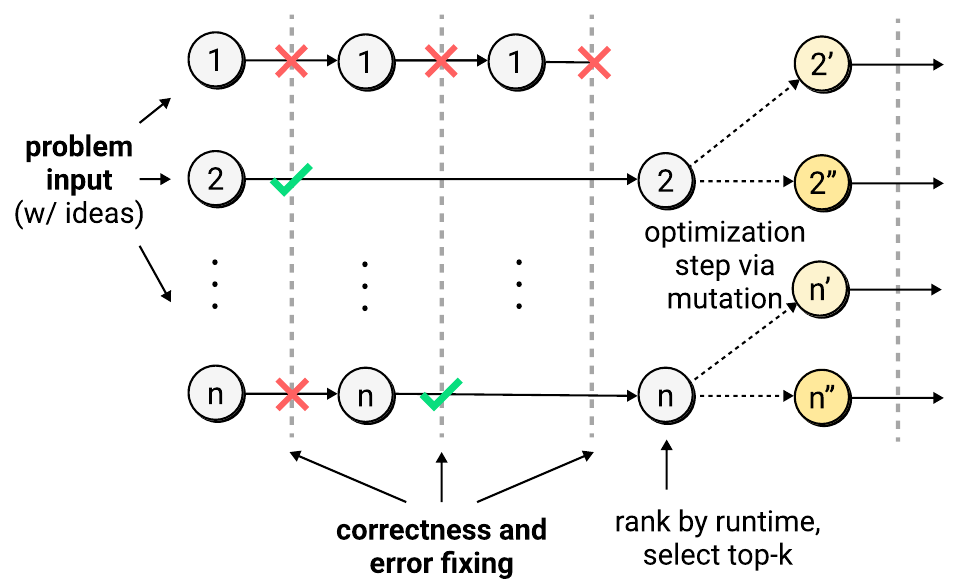}}
  \caption[PIKE-B implementation]{PIKE-B implementation, illustrating parallel evaluation and error fixing, followed by top-k selection and mutation}
  \label{fig:pike-b-impl}
\end{figure}

\begin{algorithm}
\caption{PIKE-B: Multi-Agent Branching Search with Evolutionary Mutation}
\label{alg:mas-branching}
\begin{algorithmic}
\REQUIRE Task $T$, population size $n$, error-fixing attempts allowed $m$, top candidates $k$, query budget $b$

\STATE Initialize query count $q \gets 0$
\STATE Initialize seeds $S \gets [\;]$
\STATE IBA generates $n$ initial ideas list \texttt{Ideas} using $T$

\FORALL{$I \in \texttt{Ideas}$ \textbf{in parallel}}
    \STATE Create seed $p$ with idea $I$ and code from $T$
    \STATE append $p$ to $S$
\ENDFOR

\WHILE{$q < b$}
    \FORALL{$s \in S$ \textbf{in parallel}}
        \STATE COA generates code $C$ from seed $s$
        \STATE Evaluate $C$, generates error summary $E$
        \STATE $attempts \gets 0$
        \WHILE{$E$ indicates error \AND $attempts < m$}
            \STATE ErrorFixingAgent attempts to fix $C$ using $E$
            \STATE Evaluate $\tilde{C}$, generates error summary $\tilde{E}$
            \STATE $C \gets \tilde{C}, E \gets \tilde{E}$
            \STATE $attempts \gets attempts + 1$
        \ENDWHILE
        \IF{$E$ indicates no error}
            \STATE Mark code solution as valid
        \ELSE
            \STATE Discard code
        \ENDIF
        \STATE $q \gets q + 1$
    \ENDFOR

    \STATE $S \gets [\;]$
    \STATE $Solutions \gets$ all valid solutions thus far
    \STATE Sort $Solutions$ by objective metric
    \STATE $TopK \gets$ first $k$ solutions
    \FORALL{$b \in TopK$}
        \STATE Create seed $p$ with code from $b$
        \STATE append $(n / k)$ copies of $p$ to $S$
    \ENDFOR
\ENDWHILE

\STATE \textbf{return} best overall solution based on objective metric
\end{algorithmic}
\end{algorithm}

Our PIKE-B search algorithm is outlined in Algorithm~\ref{alg:mas-branching}. We describe this strategy as an evolutionary branching search algorithm because it selects the top-k solutions found in each iteration and duplicates those solutions as seeds across the population size $n$ for the next round. This approach emphasizes exploitation, as the seed selection relies solely on the best prior solutions to guide future exploration. Specifically, the algorithm operates with a short-term memory library and does not utilize islands, resulting in a 100\% exploit-based framework. It is characterized as mutation-only evolution since all seed prompts are derived solely from a single prior solution. This makes PIKE-B particularly effective in honing in on promising solutions, leveraging previous successes to drive further optimization.

In our PIKE-B algorithm, several components of the logical framework are activated to facilitate the optimization of PyTorch models. The process begins with the library, which stores the initial model as well as valid solutions generated by the LLM. During the seed selection phase, PIKE-B selects the top-k solutions from the library, focusing exclusively on these successful prior solutions to inform the next round of optimization. The algorithm relies heavily on the evaluation stage, where generated solutions are compiled, validated, and measured based on defined performance metrics, such as speedup. If errors occur, the EFA engages to fix them, demonstrating the framework's support for iterative refinement. By leveraging these framework elements, PIKE-B aims for rapid improvements while minimizing exploration of less promising areas, effectively utilizing the structured processes designed for performance enhancement.

Figure~\ref{fig:pike-b-impl} further illustrates how seeds are selected, followed by parallel evaluation and error fixing. By focusing on a limited set of high-quality solutions, PIKE-B aims to maximize efficiency in the optimization process, making it well-suited for scenarios where rapid improvements are desired without extensive exploration of less promising areas of the space.

\subsection{PIKE-O: OpenEvolve}

OpenEvolve \cite{openevolve} is an open-source framework for LLM-based code evolution, inspired by AlphaEvolve \cite{novikov2025alphaevolve}, which enhances code through continuous evaluator feedback. To optimize code, it requires the initial code, an evaluator for validating LLM-generated solutions, and metrics to measure improvement (e.g., speedup).

Our PIKE-O implementation builds on OpenEvolve and integrates cohesively into our logical framework for PyTorch model optimization. Within OpenEvolve, the choice between short-term and long-term solution history is governed by the \textit{population} parameter, which dictates the number of solutions retained in memory, while the \textit{archive size} controls the history of elite solutions. Solutions are organized into island segments through the \textit{islands} parameter ($\geq 1$), allowing for focused exploration.

OpenEvolve features both \textit{explore} and \textit{exploit} ratio parameters; if their sum is less than 1, the remaining portion is allocated for random sampling from any segment. By default, it operates in a crossover-only mode, utilizing five or more prior elite solutions as inspirations for LLM prompts. This can be switched to mutation-only by disabling additional inspiration prompts.
In our implementation, we have modified OpenEvolve to incorporate error fixing via an EFA, filling a gap in its original capabilities for iterative refinement. Workers perform LLM querying followed by evaluation, and we enhanced these workers with the EFA loop. Notably, the IBA was not included, resulting in a purely exploit-focused mechanism during prompt generation.

Our logical framework supports parallel invocations of the optimization loop, where the seed selection phase accesses only solutions from completed iterations of prior runs. OpenEvolve's \textit{parallel evaluations} parameter manages worker distribution across islands to enable simultaneous evolution. However, this added parallelism can lead to an exploration-heavy dynamic, as workers may not wait for elite solutions to complete, potentially missing valuable exploitation opportunities. This creates a balance in the explore-exploit tradeoff, where certain configurations (e.g., disabling parallel evaluations and islands) can transition PIKE-O to an exploit-heavy approach similar to PIKE-B.

\section{Experimental Setup}

\subsection{Benchmarks: KernelBench}

The original KernelBench \cite{ouyang2025kernelbench} framework consists of 250 tasks, each comprising a self-contained PyTorch model and a random input generation function, organized into levels of increasing complexity. Level 1 includes simple operations like matrix multiplication and convolution, Level 2 combines these primitives, and Level 3 contains critical components of large ML architectures (e.g., AlexNet, MinGPT).

METR \cite{metr2025kernel} refines KernelBench by filtering tasks and adding a new Level 5. They remove Levels 1 and 2 tasks prone to noise, particularly those with low signal-to-noise ratios or that are overly simplistic. Level 4 tasks are also excluded due to their reliance on parsing complex external codebases from the HuggingFace transformers library.

We utilize the METR-refined variant of KernelBench, focusing on significant challenges by excluding Levels 1 and 2 and further filtering Level 3 to remove LSTM and GRU tasks due to high measurement noise. We refer to this filtered version as Level 3-pike, averaging 85 lines of code (LoC) excluding whitespace/comments. We also use METR’s Level 5, which includes tasks averaging 493 LoC. The largest task is HunyuanTransformer, a HunyuanVideo block with 1,268 LoC. Our final refined suite is detailed in Table~\ref{tab:metr_kbench}.

\begin{table}
\centering
\footnotesize
\caption{Refined KernelBench Suite for Evaluation}
\label{tab:metr_kbench}
\begin{tabular}{@{}p{1.25cm}p{2.25cm}p{2.0cm}c@{}}
\toprule
\textbf{Level} & \textbf{Description} & \textbf{Examples} & \textbf{Tasks} \\ 
\midrule
3-pike (Filtered) & 
Curated blocks from older models; focuses on holistic component optimization. & 
MLP, RNNs, Attention, conv. layers, Mamba components & 
30 \\ 
\addlinespace
5 (New) & 
Frontier workloads from SOTA 2024 models; focuses on novel kernel generation. & 
DeepSeek-V3, Llama 3, RWKV, SD3, Mamba-2, S4, Hunyuan Video & 
14 \\ 
\bottomrule
\end{tabular}
\end{table}

\subsection{Comparison Targets}

Numerous works have developed agent-driven systems for KernelBench benchmarks \cite{lange2025ai, lange2025towards, li2025cuda, metr2025kernel, andrews2025gpu, baronio2025kevin}. Our focus is on comparing against METR-generated solutions, as it is the only LLM-based approach tuned for H100 with publicly available outputs for our target tasks. Although the METR framework is not open-source, we ran their best solutions on our hardware.

We compare our implementations with PyTorch Eager, which uses basic kernel selection and lacks autotuning. We also compare against models compiled with TorchInductor (\verb|torch.compile|) with maximum autotuning enabled, as well as models compiled with TensorRT. We prioritize these compilers because TorchInductor has been shown to significantly outperform others with the release of PyTorch 2 \cite{pytorch2}, yet has not been compared to TensorRT.

\subsection{Code Evaluation, Correctness, and Testbed}
\label{sec:eval}

Our evaluator compiles LLM-generated solutions, verifies correctness relative to the original PyTorch model, and does performance runs. If any of these steps fail for a given solution, the sequence stops and issues/errors are passed back in the form of stdout/stderr or error tolerance issues. When all steps succeed, the final runtime is passed back. The evaluator is designed for robustness, efficiency, and accuracy. It is containerized using Docker \cite{docker}, where individual solutions are run in separate processes.

To check correctness of solutions, we adopt the same numerical equivalence tests and tolerance values as used in the prior works \cite{ouyang2025kernelbench, metr2025kernel}. We pass a randomly generated input into the original PyTorch model, and compare the output against that of the LLM-generated solution, within an error tolerance. More details can be found in Appendix~\ref{sec:correctness}.

We allocate 40+ CPU threads for our evaluator, allowing 20+ CUDA/Triton compilations to happen in parallel. All evaluations in this work are done using bare metal NVIDIA H100 GPUs with 80 GB HBM3 GPU memory connected over PCIe, as well as Intel Xeon Platinum 8480+ CPUs using allocations of 40+ CPU threads.

\subsection{Evolutionary Strategy Hyperparameters}

Each strategy utilizes all available agents unless noted otherwise. PIKE-O operates without the IBA. In both implementations, a maximum of 5 error-fixing attempts are allowed when using the EFA. We set the LLM temperature to 0.8 to encourage exploration during parallel optimization steps with identical seeds.

\textbf{PIKE-B:} After initial tuning, we select parameters: population size $n = 10$, maximum error-fix attempts $m = 5$, and top-k candidates $k = 4$. We conduct ablations on PIKE-B to analyze the effects of disabling error fixing (no EFA) and brainstorming (no IBA).

\textbf{PIKE-O:} This implementation builds on configurations from the OpenEvolve MLX Metal kernel optimization and attention optimization examples, with increased parallel evaluations. Key parameters are set as follows: population $= 25$, archive size $= 12$, islands $= 3$, explore ratio $= 0.2$, exploit ratio $= 0.7$, and parallel evaluations $= 10$.

The default PIKE-O variant differs from PIKE-B primarily in its lack of brainstorming, presence of islands, use of crossover, and support for longer-term memory. Through ablations described below, PIKE-O parameters are incrementally modified until it is virtually identical to PIKE-B, excluding brainstorming.

We experiment with various PIKE-O variants to refine its performance. The PIKE-O ({\verb|mut|}) variant narrows the focus to mutation by utilizing only one prior seed for inspiration. The PIKE-O ({\verb|mut,npar|}) variant reduces parallelism while still maintaining three islands. In contrast, the PIKE-O ({\verb|mut,npar,1isl|}) variant further simplifies the configuration by eliminating parallel evaluations and utilizing a single island. We then introduce the PIKE-O ({\verb|mut,npar,1isl,EO|}), which elevates the exploit ratio to 1, enhancing exploitation. Finally, the PIKE-O ({\verb|mut,npar,1isl,EO,SL|}) variant incorporates a short-term library with an archive size of 4 to align more closely with the PIKE-B setup. These variations allow us to explore how different configurations impact the overall performance of the PIKE-O implementations.

Due to the high cost of experimentation, we were unable to conduct a complete hyperparameter sweep for both methods within our budget. This limitation highlights the dynamic nature of our logical framework, which supports targeted optimizations while managing resource constraints effectively.

\subsection{Cost Metrics, Budget, and LLMs}

In our evaluation, we consider two related cost metrics: LLM query count and monetary cost. The LLM query count, used in prior works such as METR \cite{metr2025kernel}, is simple to track, while budget-conscious users may prioritize the monetary cost of queries.

The publicly available METR results represent the best LLM-generated code from multiple runs, each with a 300-query budget, using different LLMs like o1, Claude 3.5 Sonnet, gpt-4o, and o3-mini-high. In comparison, we use Gemini 2.5 Pro/Flash to focus on the efficacy of multi-agent techniques on a single SOTA model. Gemini 2.5 Pro is used throughout the work, with exceptions being: cheap EFA ablation (Gemini 2.5 Flash for EFA), PIKE-B using OpenAI gpt-oss-120b, and PIKE-B using OpenAI o3-mini-high. Using o3-mini-high gives us the closest comparison possible with METR, as this is the best model METR uses in their publicly available solutions.

For each benchmark run, our current PIKE implementation operates under a fixed budget of 300 queries for all agents, including EFA and IBA. This aligns with the METR budget, noting that our single 300-query runs are compared against the best results from 4 or more METR runs of 300 queries each. A 300-query run corresponds to approximately \$30-50 per task using Gemini 2.5 Pro. All tasks in our Level 3-pike experiments exceed \$25 given this budget, and \$50 for Level 5 (excluding non-Gemini runs). These amounts are comparable to the \$20 budget used in METR. All dollar amounts in this work are in USD; as of this writing, Gemini 2.5 Pro input/output prices are \$1.25/\$10.00 per 1M tokens and Gemini 2.5 Flash prices are \$0.30/\$2.50 per 1M tokens, respectively \cite{geminipricing}.

\subsection{Objective Metric: Speedup}

To measure a solution's runtime, an exclusive lock must be acquired to use a GPU. At this point, Triton's \verb|do_bench| utility is used for timing, ensuring at least 1 warmup run (which accounts for autotuning), followed by many timed runs of the benchmark, taking the mean of these runs to get a final runtime. CUDA device synchronization is performed within the timed region after the benchmark completes.

Speedup relative to PyTorch Eager is computed as \verb|eager_runtime / eval_runtime|. If speedup on any task solution is found to be $< 1$, the speedup for that solution is set to 1, given that a user could trivially fall back to using an original model with PyTorch Eager. Therefore, using our results to compare a PIKE implementation against \verb|torch.compile|, for example, one is effectively comparing against the best speedup between BOTH \verb|torch.compile| and PyTorch Eager itself.

The speedup of an approach for a given level and budget is defined as the geometric mean (geomean) of the speedups achieved on each individual task on that level, without exceeding the per-task budget for any of the tasks. For example, if we consider PIKE-B with a budget of \$25 per task on Level 3-pike, which contains 30 tasks, the total cost to run the suite under that budget amounts to \$$25 \times 30 = $ \$$750$, and the speedup is computed according to the best runtimes that PIKE-B could achieve on each task with \$25 per task.
In our evaluation, we sweep the per-task budget from zero up to the maximum budget to analyze our implementation performance trajectories as the budget increases. Geomean speedups are clamped in certain figures, meaning individual speedups $<1$ relative to Eager are clamped to speedup of 1. Speedups are clamped to reflect deployment scenarios, where a user could select between the best of PyTorch Eager and torch.compile, for example.

\subsection{Comparative Analysis}

We focus here on the complexity and granularity of the optimization steps that these LLM-based methods take. We compute the mean EFA error fix attempts required to fix broken solutions, indicating that certain strategies produce solutions which are more challenging to fix than others. To give some insight on complexity, we analyze the size of generated solutions at optimization steps by taking the SLOC (source lines of code), obtained using radon \cite{lacchia2023radon} after stripping out comment and whitespace lines.

To further understand what is happening at optimization steps, we collect a set of all \verb|(seed, generated_code)| pairs for each optimization strategy. In the case of PIKE-B, that seed is the single prior solution included in the prompt for the COA, whereas for PIKE-O the seed is the main solution that the optimizer is told to change (excluding other elite/inspiration solutions provided).

For each \verb|(seed, generated_code)| we compute the LoC (lines of code) changed at that step. This is done by adding the count of lines added and lines removed according to \verb|diff|. We then compute the mean LoC changed across steps for a task, telling us how drastically optimization steps change the code according to raw LoC. For each pair, we also use OpenAI's \verb|text-embedding-3-large| model to place each of the 2 pieces of code in an embedding space, then we measure the cosine similarity of the 2 vectors (using numpy), giving insight into the semantic similarity of generated code relative to the seed at optimization steps. We compute the mean cosine similarity across steps for a task.

\section{Results and Discussion}
\label{sec:results-discussion}

\subsection{Performance Comparison}
\label{sec:speedup-trajectories}

\begin{figure}
  \centerline{\includegraphics[width=1\columnwidth]{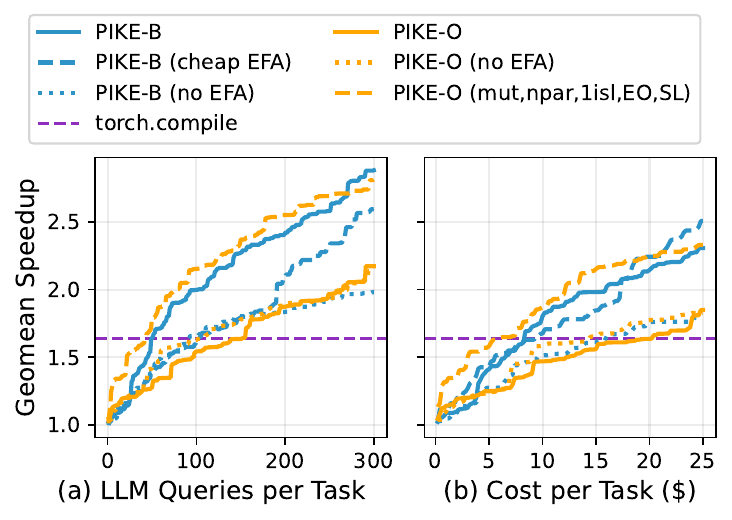}}
  \caption[Level 3-pike convergence graph with subplots]{Clamped geomean speedup across Level 3-pike tasks of each PIKE implementation by (a) LLM queries per task and (b) cost in \$ per task, all on an H100.}

  \label{fig:convergence-level3-pike}
\end{figure}

Figure \ref{fig:convergence-level3-pike}(a) illustrates the Level 3-pike geomean speedup achieved by various PIKE configurations relative to PyTorch Eager as a function of LLM queries per task. It is evident that all PIKE variants exhibit an increasing speedup with higher query counts. Notably, the PIKE-B configuration, with the EFA, demonstrates the most significant performance boost (2.88 speedup over Eager), particularly at elevated query levels. PIKE-O ({\verb|mut,npar,1isl,EO,SL|}) also shows substantial speedup (2.81 speedup over Eager), highlighting the benefits of tuning OpenEvolve into a more exploitation-focused method. Furthermore, the figure includes \verb|torch.compile| as it exhibits the best geomean speedup of all competitors. While it offers some performance enhancements, the PIKE configurations consistently outperform it, especially as the number of queries increases. PIKE-B gains substantial benefit from incorporating the EFA into the discovery loop, while default PIKE-O does not. Overall, these results underscore the effectiveness of both PIKE-B and PIKE-O in optimizing PyTorch inference.

Figure \ref{fig:convergence-level3-pike}(b) illustrates the geomean speedup achieved by different PIKE configurations based on cost per task, as opposed to the number of LLM queries. This analysis highlights the tradeoffs between performance gains and cost, demonstrating how each configuration scales with budget constraints. Notably, the PIKE-B configuration with a cheap EFA (Gemini 2.5 Flash) provides the most significant speedup of 2.51 for a \$25 per task budget, indicating this method offers the best performance return on investment. Close behind are PIKE-B and PIKE-O ({\verb|mut,npar,1isl,EO,SL|}) with the expensive EFA (Gemini 2.5 Pro), achieving speedups of 2.31 and 2.33 for the same budget, respectively. Additionally, while \verb|torch.compile| is a viable option, especially for budgets below \$10 per task, it falls short of the performance offered by the PIKE methods, particularly as costs rise. Overall, considering both cost and performance, PIKE-B (especially with cheap EFA) stands out as the optimal choice for maximizing inference efficiency while effectively managing expenses.

\begin{figure}[t]
  \centerline{\includegraphics[width=1\columnwidth]{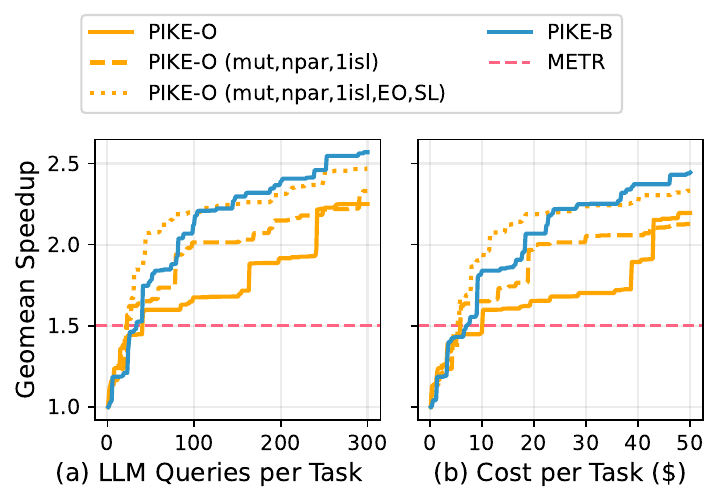}}
  \caption[Level 5 convergence graph with subplots]{Clamped geomean speedup across Level 5 tasks of each PIKE implementation by (a) LLM queries per task and (b) cost in \$ per task, all on an H100.}

  \label{fig:convergence-level5}
\end{figure}

\begin{figure}
\centering

\subfigure[Level 3-pike, dashed line shows budget \$25]{
\includegraphics[width=0.9\columnwidth]{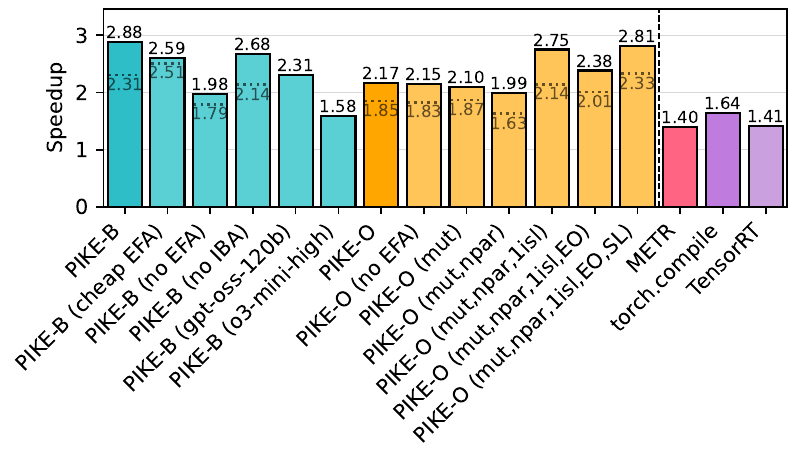}
\label{fig:speedup-level3-pike}
}

\vspace{1em}

\subfigure[Level 5, dashed line shows budget \$50]{
\includegraphics[width=0.72\columnwidth]{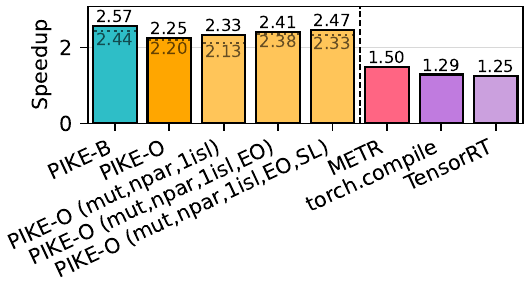}
\label{fig:speedup-level5}
}

\caption[Overall speedups]{Clamped speedup relative to Eager of our approaches (including extra ablations) and other approaches, using an H100. All of our approaches get a budget of exactly 300 LLM queries.}
\label{fig:overall-speedups}
\end{figure}

Figure \ref{fig:convergence-level5} presents a comparative analysis of geomean speedup achieved by different PIKE configurations across the Level 5 benchmark suite, focusing on LLM queries per task (Figure \ref{fig:convergence-level5}(a)) and cost per task (Figure \ref{fig:convergence-level5}(b)). In terms of speedup versus the number of LLM queries, PIKE-B shows significant improvements over PIKE-O as the query count increases (with speedups of 2.57 over 2.25, respectively), maintaining a clear advantage throughout. For speedup by cost per task, PIKE-B again outperforms PIKE-O, although both configurations provide respectable gains as costs rise (with respective speedups of 2.44 and 2.2 for a \$50 per task budget). These results indicate that while both configurations enhance performance relative to budget, PIKE-B remains the leading choice, demonstrating consistently higher geomean speedup across varying costs.

However, the exploitation-tuned PIKE-O ({\verb|mut,npar,1isl,EO,SL|}) achieves speedup of 2.47 with 300 queries per task budget, and 2.33 with \$50 per task budget, reaching quite similar performance levels to PIKE-B. This is a notable improvement over PIKE-O ({\verb|mut,npar,1isl|}) which achieves speedup 2.33 with a 300 query budget and 2.13 with a \$50 per task budget. This improvement suggests that library size may play a role in the advantage of PIKE-O ({\verb|mut,npar,1isl,EO,SL|}) over default PIKE-O, as we explore further in \S\ref{sec:exploitation-oe}.

METR is included for comparison due to having the best geomean speedup of all competitors. A METR trajectory is not shown because only final METR solutions are available publicly. It shows lower speedup (1.5 for the best solutions across tasks), underscoring the effectiveness of the PIKE approaches in optimizing efficiency for Level 5 tasks. Overall, these findings affirm the advantage of PIKE-B in maximizing inference efficiency and performance, making it the optimal choice across benchmark tasks and costs.

\subsection{Component \& Comparative Analysis}

\subsubsection{PIKE-B Ablation}
\label{sec:pike-b-ablation}

In this section, we analyze the various configurations of PIKE-B to understand their individual contributions to performance improvements. The algorithm employs an evolutionary strategy which is exploit-heavy and mutation-based, with results from the Level 3-pike suite, shown in Figure \ref{fig:speedup-level3-pike}, indicating distinct performance characteristics for each configuration.

The PIKE-B (with expensive EFA) configuration achieves the highest speedup of 2.88, demonstrating that robust error correction is crucial for optimizing system performance. The EFA effectively identifies and rectifies errors in the code generated by the COA, ensuring high-quality outputs. In contrast, the PIKE-B (no EFA) variant shows diminished performance with a speedup of only 1.98. The absence of an error correction mechanism significantly decreases its effectiveness, highlighting the EFA's critical role in maintaining optimization efficiency.

The PIKE-B (no IBA) configuration, with a speedup of 2.68, illustrates that while brainstorming generates ideas, its impact on performance is comparable to configurations using the IBA, which achieve speedups of 2.88 (PIKE-B) and 2.59 (PIKE-B with cheap EFA). The PIKE-B (cheap EFA) configuration produces a speedup of 2.59, balancing cost and performance but still relying on error correction to avoid pitfalls in generating untested optimizations. Using cheap EFA brings the mean cost per EFA query across all Level 3-pike tasks from \$0.15 to \$0.04 for PIKE-B, but this comes at the expense of error fixing success, as the percentage of solutions working before the 5 attempt EFA max drops from 79.3\% for original PIKE-B to 71.1\% with cheap EFA (see \S\ref{sec:performance-gaps} for details on PIKE-B statistics).

The PIKE-B (o3-mini-high) configuration, with a speedup of 1.58, confirms that PIKE-B outperforms METR (with speedup 1.40) when using identical models (o3-mini-high), the same query budget (300), and the same target hardware (H100). To be precise, METR performs best-of-N between o3-mini results and other (worse performing) models, each with a 300 query budget. This means the PIKE-B approach is still strictly better, since this ablation only uses o3-mini-high. The PIKE-B (gpt-oss-120b) configuration, with a speedup of 2.31, shows that PIKE-B is still quite effective when combined with a less powerful open-weight model.

These insights suggest that an optimal setup for developing effective LLM-based multi-agent systems for PyTorch inference should emphasize an exploitative approach that leverages the best-known solutions while incorporating mutations to explore variations. A critical recommendation is to prioritize resources toward robust error correction mechanisms through the EFA to ensure reliable optimizations. In conclusion, focusing on error correction, particularly through the EFA, while utilizing strategies that exploit high-quality solutions will lead to a more robust, efficient, and high-performing system.

\subsubsection{PIKE-O Ablation}
\label{sec:exploitation-oe}

To understand the performance of PIKE-O, we evaluate several configurations under the Level 3-pike and Level 5 tasks. The default PIKE-O is designed with three independent islands, each maintaining a recent-program buffer and long-term library of solutions (elite archive) for exploitation sampling, while enabling parallel evaluations. While structurally similar to PIKE-B in terms of general evolutionary principles, the default PIKE-O differs in its distribution of search effort across multiple islands and its default parallelism level. Due to time and cost constraints, we do not evaluate all combinatorial versions of PIKE-O components. We instead focus on consecutive steps which align PIKE-O more closely with PIKE-B in our logical framework.

The Level 3-pike results are shown in Figure \ref{fig:speedup-level3-pike}. The base PIKE-O configuration for Level 3-pike achieves a speedup of 2.17 with EFA enabled. Removing EFA yields a nearly identical speedup of 2.15, indicating that, unlike PIKE-B, the base PIKE-O’s performance is not substantially affected by error correction. This suggests that its multi-island architecture and inherent redundancy in seed diversity compensate for occasional errors, reducing EFA’s marginal benefit.

The PIKE-O ({\verb|mut|}) variant modifies the prompting behavior. The basic PIKE-O resembles a crossover-heavy evolutionary strategy, often including 5 or more prior elite or diverse solutions as inspiration alongside a target seed program for modification. In PIKE-O ({\verb|mut|}), we deliberately remove these additional inspirations and provide exactly one previous seed program per prompt, making the search mutation‑focused and directly comparable to PIKE-B’s mutation‑only mode. This adjustment leaves performance virtually unchanged from PIKE-O, with a speedup of 2.10.

PIKE-O ({\verb|mut,npar|}), which reduces parallelism but retains three islands, drops to 1.99. Multiple islands keep the strategy exploration-oriented despite the parallelism constraint, reducing the exploitative nature that benefits the best-performing variant. This difference also highlights the unintended exploratory bias introduced by high parallelism (\textit{parallel evaluations} = 10): quick-to-evaluate but low-quality solutions can push the algorithm to generate more early-stage seeds rather than maturing complex solutions that achieve better performance.

The PIKE-O ({\verb|mut,npar,1isl|}) variant demonstrates a performance boost, achieving a speedup of 2.75. In this configuration, the number of islands is set to one. Building on this, the most notable result is observed in PIKE-O ({\verb|mut,npar,1isl,EO,SL|}), which attains the highest speedup of 2.81. This variant eliminates parallel evaluations (setting to 1), reduces the number of islands from three to one, sets the exploit ratio to 1, and utilizes just 4 elites in the short-term library. These changes sharply limit exploration and shift the focus toward an exploit-heavy approach.

In contrast, the PIKE-O ({\verb|mut,npar,1isl,EO|}) variant, which also has a high exploit ratio, achieves only a modest speedup of 2.38. This suggests that the speedup difference between PIKE-O ({\verb|mut,npar,1isl|}) and PIKE-O ({\verb|mut,npar,1isl,EO,SL|}) is largely due to the limited size of the short-term library in the latter.

The operational mode of PIKE-O ({\verb|mut,npar,1isl,EO,SL|}) closely mirrors that of PIKE-B’s mutation-exploitation strategy. Both methods perform similarly, achieving approximate speedups of 2.3 at a \$25 budget. In this setup, each iteration builds directly upon the most promising recent solution, thereby maximizing the payoff from proven code paths.

The Level 5 results are illustrated in Figure \ref{fig:speedup-level5}. The default PIKE-O variant shows the lowest performance, achieving a speedup of 2.25 at full budget. In contrast, PIKE-O ({\verb|mut,npar,1isl,EO,SL|}) emerges as the top PIKE-O performer, recording a speedup of 2.47 for the full budget and 2.33 for the \$50 per task budget. The differences in performance among the variants are less pronounced than in Level 3-pike; for instance, the second-best variant, PIKE-O ({\verb|mut,npar,1isl,EO|}), attains a speedup of 2.41 at full budget and outperforms the former variant for the \$50 budget. Closely following is PIKE-O ({\verb|mut,npar,1isl|}), which lags behind with speedups of 2.33 for the full budget and 2.13 for the \$50 per task budget.

The difference between the exploitation-focused PIKE-O variants is less pronounced in Level 5. However, the main takeaway is that, compared to the default PIKE-O variant, which combines exploitative and exploratory strategies, the more exploitative configurations yield better performance for this type of task. Notably, while PIKE-O ({\verb|mut,npar,1isl,EO,SL|}) consistently remains the top performer, the second and third places have switched, with PIKE-O ({\verb|mut,npar,1isl,EO|}) outpacing PIKE-O ({\verb|mut,npar,1isl|}). This shift suggests that maximizing the exploit ratio can be beneficial.

Overall, these results indicate that the greatest improvements in PIKE-O performance arise from adjusting the search dynamics towards exploitation. By reducing the number of islands, limiting parallelism, and minimizing the elite archive, we achieve a focused, mutation-driven optimization that closely aligns with the strengths observed in PIKE-B.

\subsubsection{Competitors}

Alternative solutions such as \verb|torch.compile|, TensorRT, and METR for Level 3-pike (Figure~\ref{fig:speedup-level3-pike}) show diminished speedup performance, recorded at 1.64, 1.41, and 1.40, respectively. Similarly diminished performance is demonstrated in Level 5 (Figure~\ref{fig:speedup-level5}), with the methods achieving speedups of 1.29, 1.25, and 1.50, respectively. This indicates that these alternatives do not optimize as well as the PIKE methodologies under the budgets used. Overall, the results further illustrate the effectiveness of the PIKE approaches in enhancing performance, particularly highlighting the PIKE-B and exploitation-tuned PIKE-O configurations as the most beneficial options for users pursuing optimal speedups in their applications.

\subsubsection{Dynamics of PIKE-B and PIKE-O}
\label{sec:performance-gaps}

\begin{figure}
\centering

\begin{minipage}[b]{0.8\columnwidth}
    \includegraphics[width=\textwidth]{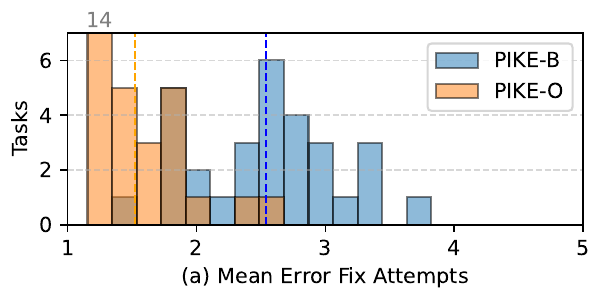}
\end{minipage}
\begin{minipage}[b]{0.8\columnwidth}
    \includegraphics[width=\textwidth]{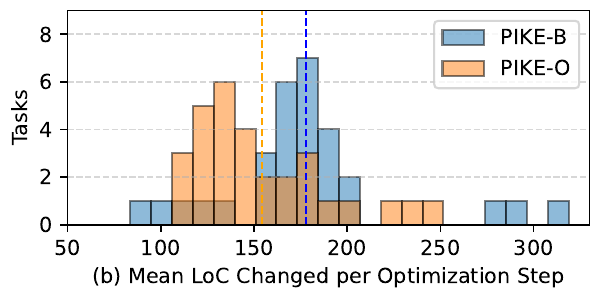}
\end{minipage}

\caption[Level 3-pike PIKE statistics]{PIKE-B and PIKE-O Level 3-pike per step analysis of (a) correctness attempt count (b) LoC changed. Dashed lines indicate mean of means for each implementation.}
\label{fig:hists-1-pikeo-pikeb}
\end{figure}

For the PIKE-B and PIKE-O configurations, the mean number of error-fix attempts per task was computed and binned into the histogram as shown in Figure~\ref{fig:hists-1-pikeo-pikeb}(a). PIKE-O requires fewer fixes, with most tasks averaging 1.5 attempts or less, while PIKE-B generally exceeds 2 attempts, indicating more challenging repairs. Since EFA prompts are identical for both methods, this difference arises from their distinct seeding and prompting strategies. Figure~\ref{fig:hists-1-pikeo-pikeb}(b) further shows that PIKE-B modifies more LoC per optimization step than PIKE-O. Taken together, the data show that within the 300-query budget, exploit-heavy PIKE-B performs larger, riskier code transformations, whereas PIKE-O makes smaller, more conservative changes that are easier to correct.

\begin{figure}
\centering

\begin{minipage}[b]{0.8\columnwidth}
    \includegraphics[width=\textwidth]{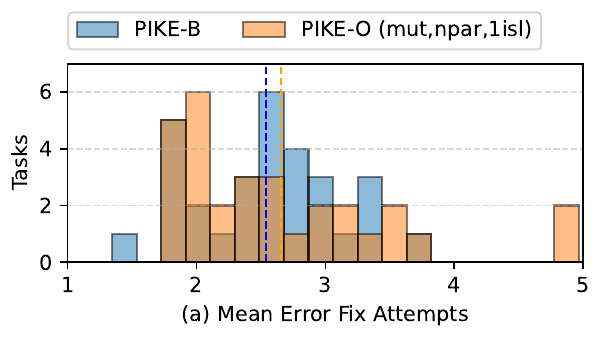}
\end{minipage}
\begin{minipage}[b]{0.8\columnwidth}
    \includegraphics[width=\textwidth]{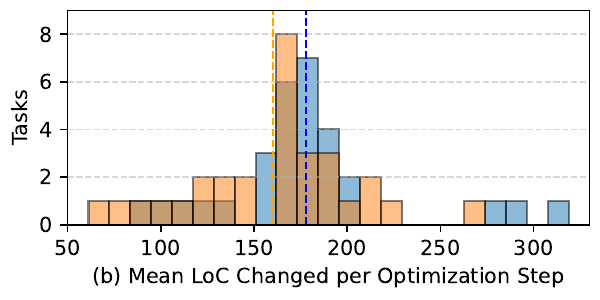}
\end{minipage}

\caption[Level 3-pike PIKE statistics]{PIKE-B and PIKE-O \texttt{(mut,npar,1isl)} Level 3-pike per-step analysis of (a) correctness attempt count (b) LoC changed. Legend is shared between (a) and (b). Dashed lines indicate mean of means for each implementation.}
\label{fig:hists-2-pikeobest-pikeb}
\end{figure}

We also compare PIKE-B with the exploitative PIKE-O variant ({\verb|mut,npar,1isl|}), as the PIKE-O ablation study shows that tuning toward heavier exploitation yields performance comparable to PIKE-B. 
On Level~3-pike, Figure~\ref{fig:hists-2-pikeobest-pikeb}(a) indicates that this PIKE-O variant requires slightly more error-fix attempts than PIKE-B, with some tasks exceeding four attempts. Figure~\ref{fig:hists-2-pikeobest-pikeb}(b) shows that its mean lines of code changed per optimization step are slightly lower than PIKE-B. 
Together, these results suggest that PIKE-B and exploitative PIKE-O behave similarly when considering both error-fix effort and the magnitude of code changes per step.

We did not include PIKE-O ({\verb|mut,npar,1isl|}) without EFA due to budget constraints. We infer that it would struggle without EFA, given that solutions show similar error-fixing behavior to PIKE-B. Therefore, we conjecture that the need for error-fixing is driven by exploit-heavy optimization which rapidly increases code complexity.

\begin{table}[b]

\caption{Statistics for PIKE-B and PIKE-O on Level 3-pike}
\label{tab:statistics}

\centering
\footnotesize
\begin{tabular}{@{}p{3.9cm}p{1.2cm}p{1.2cm}@{}}
\toprule
Statistic & PIKE-B & PIKE-O \\
\midrule
Mean SLOC of generated code at optimization steps & 244 & 169 \\
\addlinespace
Mean steps completed per task with budget \$25 & 160 & 198 \\
\addlinespace
Mean cost per task with LLM query budget of 300 & \$50.96 & \$39.59 \\
\addlinespace
Percent solutions working before EFA 5 attempt max & 79.3\% & 96.7\% \\
\addlinespace
Mean cos sim. between steps & 0.91 & 0.87 \\
\addlinespace
Best sols. using CUDA / Triton & 6 / 23 & 19 / 11 \\
\bottomrule
\end{tabular}

\end{table}

Additional performance statistics for PIKE-B and PIKE-O are presented in Table~\ref{tab:statistics}. PIKE-B tends to generate larger programs during optimization steps (mean SLOC~=~244 vs.~169 for PIKE-O). As a result, fewer optimization steps can be completed within the fixed task budget of~\$25 (160 for PIKE-B vs.~198 for PIKE-O), and the overall cost to reach a fixed LLM query budget of 300 is higher (\$50.96 vs.~\$39.59).  

Error correction outcomes also differ. PIKE-O’s solutions can more frequently be made correct within the five-attempt EFA budget (96.7\% vs.~79.3\%), suggesting that its exploratory bias combined with smaller code outputs leads to fewer syntactic or semantic errors at generation time. In both methods, the EFA succeeds in repairing most invalid programs within the five-attempt limit, though these results indicate that further tuning may be beneficial for challenging tasks or more exploit-heavy strategies.  

Despite substantial LoC changes during optimization, the semantic similarity between seeds and optimized solutions remains high in both systems (mean cosine similarity~=~0.91 for PIKE-B,~0.87 for PIKE-O). This indicates that optimizations often preserve the semantic structure of seeds while introducing targeted changes. 

A particularly interesting trend is observed in the distribution of best solutions across backends. The better-performing, exploit-heavy PIKE-B produces more top solutions in Triton (23 vs.~11), whereas the more 
explore-oriented PIKE-O yields more top solutions in CUDA (19 vs.~6). This may reflect differing backend advantages aligned to each search style: exploit-heavy strategies can better leverage Triton’s autotuning 
capabilities, while exploration tends to uncover more CUDA-specific optimizations.

\subsection{Task-wise Benchmark Analysis}
\label{sec:taskwise-results}

Speedup breakdowns for primary PIKE implementations and competitors on Level 3-pike and Level 5 with budget 300 queries are shown in Table~\ref{tab:breakdown-level3-pike} and Table~\ref{tab:breakdown-level5}, respectively. Best speedups for a given task are bolded. Clamped geomean sets any individual speedups $<1$ to 1, as any of the approaches could trivially fall back to PyTorch Eager in a deployment setting. Success rates of each approach are shown, along with number of tasks slower than Eager and torch.compile (TI).

Geomean relative to torch.compile is shown as well, which only includes tasks where torch.compile succeeds. PIKE strategies suffer on Level 5 with this metric, due to PIKE showing good speedup on many tasks where torch.compile fails entirely. PIKE strategies perform better on geomean (eager/TI) metric for Level 5, given that tasks where torch.compile fails simply fall back to PyTorch Eager as the baseline runtime.

Across both Level~3-pike and Level~5, the largest speedups arise from a small set of recurring optimization patterns. The most impactful is precision reduction (e.g., FP16) combined with replacing default PyTorch attention with custom Triton flash attention kernels, often enhanced with fused projection operations and backend autotuning. Full-model or large subgraph fusion into a single custom CUDA/Triton kernel is another consistent winner, eliminating intermediate memory writes and reducing launch overhead. Reordering expensive compute and pooling operations to reduce input size before heavy compute, as well as targeted operator fusion within monolithic Triton kernels, also contribute to large gains. These strategies often stack, yielding double-digit speedups on challenging models.

Furthermore, a small number of tasks were removed due to pathological baselines (e.g., LSTM/GRU with extreme launch overhead), and some invalid or noisy results were excluded from aggregates. Overall, the results suggest that precision reduction, aggressive kernel fusion, and architecture‑specific tuning form a general recipe for substantial performance improvements across tasks, with similar patterns emerging in both PIKE and the strongest competing systems. A detailed analysis of all of the performance outliers and their strategies can be found in the Appendix~\ref{app:task-analysis}.

\begin{table}[H]

\caption{Breakdown of speedups on Level 3-pike relative to Eager. PB = PIKE-B, PO' = PIKE-O \texttt{(mut,npar,1isl,EO,SL)}, TI = TorchInductor (torch.compile), MET = METR, TRT = TensorRT. gmean (TI) = unclamped geomean w.r.t. torch.compile, gmean (eager/TI) = unclamped geomean w.r.t. best of Eager or torch.compile}
\label{tab:breakdown-level3-pike}

\centering
\small
\setlength{\tabcolsep}{4pt}

\begin{tabular}{lrrrrr}
\toprule
Task & PB & PO' & MET & TI & TRT \\
\midrule
VanillaRNN & 0.93 & \textbf{1.05} & \textbf{1.05} & {—} & {—} \\
SwinMLP & 1.03 & 1.50 & 1.13 & 1.46 & \textbf{1.69} \\
DenseNet121 & 1.16 & \textbf{1.63} & 0.61 & \textbf{1.63} & 0.64 \\
MLP & \textbf{1.20} & 1.00 & 0.96 & 1.12 & 1.18 \\
MobileNetV2 & 1.23 & \textbf{3.35} & 0.96 & 0.87 & 0.58 \\
SwinV2 & 1.23 & 1.59 & 1.19 & 1.92 & \textbf{2.44} \\
ResNet101 & 1.25 & \textbf{4.12} & {—} & 0.65 & 0.50 \\
ShallowWideMLP & 1.33 & \textbf{1.41} & 1.06 & 1.29 & 1.24 \\
EfficientNetB2 & 1.36 & 1.35 & {—} & \textbf{2.86} & 0.98 \\
GoogleNetIM & 1.55 & 1.00 & 1.51 & 0.95 & \textbf{1.69} \\
Mamba2ReturnY & 1.79 & 2.29 & 0.44 & \textbf{4.58} & {—} \\
ShuffleNet & 2.01 & \textbf{2.70} & 1.57 & 0.91 & 1.05 \\
DenseNet201 & \textbf{2.40} & 1.13 & 0.71 & 1.33 & {—} \\
LeNet5 & \textbf{2.58} & 1.68 & 1.16 & 1.45 & 1.19 \\
RegNet & 2.58 & \textbf{3.51} & 0.26 & 2.19 & 1.49 \\
EfficientNetM & 2.61 & \textbf{3.51} & {—} & 0.89 & 1.06 \\
ResNet18 & 2.99 & \textbf{3.45} & 1.04 & 2.15 & 0.62 \\
ResNetBB & \textbf{3.00} & 2.55 & 2.19 & 1.91 & 1.50 \\
MobileNetV1 & \textbf{3.39} & 1.24 & 1.22 & 1.36 & 0.57 \\
DenseNet121DB & \textbf{3.41} & 1.44 & 1.22 & 2.05 & 1.45 \\
ReLUSelfAtt & 3.61 & \textbf{3.87} & {—} & 1.87 & {—} \\
EfficientNetB1 & \textbf{3.70} & 1.37 & 1.33 & 1.47 & 0.58 \\
EfficientNetB0 & \textbf{4.64} & 1.91 & 1.40 & 1.68 & 0.58 \\
MinGPTBlock & 4.73 & 10.91 & \textbf{11.83} & 1.26 & 4.92 \\
ShuffleNetUnit & \textbf{5.48} & 4.33 & 1.86 & 1.62 & 2.76 \\
SqueezeNet & 6.50 & \textbf{8.56} & 1.22 & 4.16 & 4.24 \\
MinGPTCausAtt & 10.83 & \textbf{14.59} & 12.13 & 1.25 & 3.93 \\
Mamba2ReturnF & 12.26 & \textbf{15.01} & 1.11 & 5.48 & {—} \\
DenseNet121TL & 12.98 & \textbf{13.00} & 2.75 & 2.48 & 2.65 \\
VisionAtt & \textbf{28.67} & 10.34 & 1.00 & 0.92 & 1.10 \\
\midrule
gmean (clamped) & \textbf{2.88} & 2.81 & 1.40 & 1.64 & 1.41 \\
gmean (uncl.) & \textbf{2.87} & 2.81 & 1.27 & 1.59 & 1.24 \\
gmean (TI) & \textbf{1.85} & 1.80 & 0.80 & 1.00 & 0.87 \\
gmean (eager/TI) & \textbf{1.75} & 1.72 & 0.79 & 0.97 & 0.84 \\
success & 100\% & 100\% & 87\% & 97\% & 83\% \\
slower & 1 & 1 & 6 & 6 & 8 \\
slower (TI) & 5 & 8 & 16 & 0 & 13 \\
\bottomrule
\end{tabular}

\end{table}

\begin{table}

\caption{Breakdown of speedups on Level 5 relative to Eager. PB = PIKE-B, PO' = PIKE-O \texttt{(mut,npar,1isl,EO,SL)}, TI = TorchInductor (torch.compile), MET = METR, TRT = TensorRT. gmean (TI) = unclamped geomean w.r.t. torch.compile, gmean (eager/TI) = unclamped geomean w.r.t. best of Eager or torch.compile}
\label{tab:breakdown-level5}

\centering
\small
\setlength{\tabcolsep}{4pt}

\begin{tabular}{lrrrrr}
\toprule
Task & PB & PO' & MET & TI & TRT \\
\midrule
DeepSeek3 & 1.00 & \textbf{1.01} & 1.00 & {—} & {—} \\
RWKVTorch & 1.00 & 0.87 & 1.00 & \textbf{2.35} & 0.83 \\
Llama2Dec & 1.18 & \textbf{2.31} & {—} & 1.09 & {—} \\
Llama2 & 1.28 & \textbf{1.40} & {—} & {—} & {—} \\
StableDiff3 & 1.29 & 2.23 & \textbf{5.75} & 1.11 & 3.87 \\
DeepSeek3MLA & 1.30 & \textbf{1.31} & 1.00 & 1.04 & {—} \\
HunyuanTrans & \textbf{1.46} & {—} & 0.99 & {—} & {—} \\
DeepSeek3MLAD & \textbf{1.88} & 1.02 & {—} & 1.13 & {—} \\
DeepSeek3MOEl & 3.06 & \textbf{4.29} & 1.05 & 0.94 & {—} \\
S4 & 4.25 & 2.72 & \textbf{9.89} & 0.85 & {—} \\
HunyuanEnc & 6.77 & \textbf{9.74} & 1.00 & {—} & 2.72 \\
HunyuanDec & \textbf{8.72} & 4.02 & 1.00 & {—} & 2.09 \\
DeepSeek3MOEs & 9.41 & \textbf{10.38} & {—} & {—} & {—} \\
Mamba2 & \textbf{10.81} & 6.78 & 4.99 & 10.57 & 0.70 \\
\midrule
gmean (clamped) & \textbf{2.57} & 2.47 & 1.50 & 1.29 & 1.25 \\
gmean (uncl.) & \textbf{2.57} & 2.44 & 1.50 & 1.27 & 1.20 \\
gmean (TI) & 1.44 & 1.42 & \textbf{2.10} & 1.00 & 0.43 \\
gmean (eager/TI) & \textbf{1.99} & \textbf{1.99} & 1.52 & 0.97 & 0.86 \\
success & 100\% & 93\% & 71\% & 57\% & 36\% \\
slower & 2 & 1 & 3 & 2 & 2 \\
slower (TI) & 1 & 3 & 2 & 0 & 2 \\
\bottomrule
\end{tabular}

\end{table}

\section{Related Work}

Agent-based systems range from single-agent designs for sequential, context-rich tasks, to multi-agent systems (MAS) that split work among specialized agents for profiling, planning, coding, and testing. Mixture-of-Agents approaches combine outputs from different LLMs (commercial or open-source) into improved results, as shown by AutoGen \cite{wu2024autogen} and MetaGPT \cite{hong2024metagpt}. Benchmarks like Terminal-Bench \cite{tbench_2025}, Recovery-Bench \cite{tan2025recovery}, SWE-bench \cite{jimenez2023swe}, GSO \cite{shetty2025gso}, BuildBench \cite{zhang2025buildbench}, and MLAgentBench \cite{huang2024mlagentbench} reveal that architecture often matters more than the base LLM, especially in LLM-based optimization, which needs long-horizon planning and integration with tools. Evolutionary MAS variants such as AlphaEvolve \cite{novikov2025alphaevolve}, OpenEvolve \cite{openevolve}, GEPA \cite{agrawal2025gepa}, AgentNet \cite{yang2025agentnet}, EvoAgent \cite{yuan2024evoagent}, SE-agent \cite{lin2025se}, and LEAR \cite{gurkan2025lear} demonstrate iterative search and adaptation for code, model, and prompt optimization. Cost-focused strategies like FrugalGPT \cite{chen2024frugalgpt}, Yue et al. \cite{yue2024large}, and BudgetMLAgent \cite{gandhi2024budgetmlagent} cascade cheaper models for reasoning tasks, but budget-aware MAS comparisons remain unexplored.

In GPU code optimization, model-level compilers such as TorchInductor \cite{pytorch2}, TensorRT \cite{tensorrt}, ONNX Runtime \cite{onnxruntime}, and XLA \cite{xla} automate graph optimizations, kernel fusion, and selection from libraries like cuBLAS/cuDNN \cite{cublas, chetlur2014cudnn}, or generate kernels via libraries/DSLs like Triton \cite{tillet2019triton}, CUTLASS \cite{kerr2017cutlass}, and ThunderKittens \cite{spector2024thunderkittens}. Autotuners like Kernel Tuner \cite{van2019kernel}, CLTune \cite{nugteren2015cltune}, and ATF \cite{rasch2018atf} empirically search kernel parameters using heuristics or metaheuristics, reducing manual engineering while exploiting complex hardware.

Agent-Driven PyTorch/CUDA Optimization replaces traditional compiler/autotuner workflows with LLM agent pipelines for end-to-end GPU kernel engineering. Lange et al. \cite{lange2025towards, lange2025ai} translate PyTorch to CUDA, iteratively optimizing via evolutionary search, ensembling, and LLM-based verification to avoid incorrect solutions—highlighting KernelBench Level 1–2 issues. Li et al. \cite{li2025cuda} introduce CUDA-L1, combining fine-tuning, self-supervised learning, and contrastive RL with runtime feedback for generalizable optimizations and reward-hacking prevention. METR  \cite{metr2025kernel} refine KernelBench by removing flawed tasks and adding Level 5 workloads, achieving strong results with KernelAgent yet still trailing expert engineers. Andrews et al. \cite{andrews2025gpu} present GPU Kernel Scientist, iterating AMD kernels through experiment-driven loops without profilers. Chen et al. \cite{chen2025cuda} propose Feature Search and Reinforcement (FSR) for validated and profiled architecture-aware speedups. Baronio et al. \cite{baronio2025kevin} train Kevin, a multi-turn RL CUDA optimizer balancing correctness and performance, exceeding frontier models. Wei et al. \cite{wei2025astra} offer Astra, a MAS targeting production SGLang \cite{zheng2025sglang} kernels, coordinating generation, testing, profiling, and planning agents for real-world speedups.

\section{Future Work}

One promising direction is to extend the framework to support multi-GPU and distributed inference workloads, broadening applicability to production-scale deployment scenarios. It is also worth exploring adaptive strategies that dynamically shift between exploration and exploitation during the search, or use early cutoff mechanisms. Another promising direction is to enable layout algebra generation (e.g. CuTe layouts), mirroring the manual performance engineering process used when developing CUTLASS kernels.

\section{Conclusion}

Maximizing AI model performance on target GPU hardware remains an ongoing problem, and numerous traditional and LLM-based solutions have been proposed. We have presented a logical framework for comparing multi-agent PyTorch optimization systems. We develop and evaluate a range of implementations that fit into this framework. Our best implementation achieves an average speedup of 2.88× over PyTorch Eager (1.85× over torch.compile)  on an H100 GPU over a range of complex KernelBench tasks. We find that exploit-heavy optimization strategies paired with an error-fixing agent perform best. We find strategy performance to relate to optimization step granularity, with more substantial steps leading to better performance.

\section*{Acknowledgements}

This material is based upon work supported by the U.S. Department of Energy, Office of Science, Office of Advanced Scientific Computing Research, through ``Advancements in Artificial Intelligence for Science'', DE-FOA-0003264 award number FP00018807, under contract number DE-AC02-05CH11231.
Additionally, this paper used resources of the National Energy Research Scientific Computing Center, which is supported by the Office of Science of the U.S. Department of Energy under Contract No. DE-AC02-05CH11231.
The authors would like to thank the Information Technology Division at Lawrence Berkeley National Laboratory (LBL-IT) for their support. This work benefited from the assistance of CBorg Chat, an AI-powered chat assistant developed by LBL-IT.
Kirill Nagaitsev was supported by the U.S. Department of Energy Computational
Science Graduate Fellowship (DE-SC0024386).

\bibliography{bib/paper}
\bibliographystyle{mlsys2025}

\appendix

\section{Correctness}
\label{sec:correctness}

To check correctness of solutions, we adopt the same numerical equivalence tests and tolerance values as used in the prior works. More specifically, each benchmark provides a function for generating random inputs of correct dimensions for the problem. We pass this input into the original PyTorch model, using the output as the gold standard output to compare against. We pass the input into LLM-generated solutions as well. Then, we compare to the gold standard output using \verb|torch.allclose| \cite{torchallclose} with \verb|atol = 0.01| (absolute tolerance) and \verb|rtol = 0.01| (relative tolerance), exactly as in prior works \cite{ouyang2025kernelbench, metr2025kernel}. A solution is considered valid if it satisfies this element-wise condition, where \textit{gold} is the gold standard output and \textit{other} is the output of the LLM-generated solution:

\[
| \text{gold}_i - \text{other}_i | \leq \verb|atol| + \verb|rtol| \times |\text{other}_i|
\]

Using an error tolerance is needed to account for floating point imprecision between models. As a result, this also allows precision reduction to FP16 on certain problems.

\subsection{Numerical Safety}

METR uses precision reduction in its best available solutions, so we allow precision reduction to make a fair comparison against METR’s LLM-generated solutions. Safely applying precision reduction to ML models is challenging in practice, and existing approaches such as NVIDIA Model Optimizer’s AutoCast \cite{nvidiamodelopt} require calibration datasets to analyze tensor magnitudes and ensure overflow/underflow are avoided. Precision reduction is safe for the sets of weights and inputs we test, but making stronger guarantees would require further analysis.

To understand the impact of precision reduction on the performance that PIKE achieves, we try removing all solutions that involve precision reduction. For PIKE-B on Level 3-pike, the best solutions from the remaining pool result in overall speedup of 2.12 over PyTorch eager (down from 2.88, but still substantially outperforming torch.compile). Performance without precision reduction could likely improve if prompts explicitly discouraged it, though our current prompts do not.

\section{Variability \& Robustness}

We conducted two extra end-to-end runs of PIKE-B on Level 3-pike, in order to investigate the robustness of the method. This gave a total of 3 PIKE-B runs for Level 3-pike, with geomeans over PyTorch Eager of 2.76, 2.88, and 2.89. For each individual task, we took the median of its 3 runtimes, giving us a median speedup for each task. We then computed a geomean speedup of 2.83 over PyTorch Eager from these medians, negligibly lower than our headline speedup of 2.88 for PIKE-B. Though limited by budget constraints from doing further variability analysis, this helps demonstrate that overall performance is relatively consistent across runs.

\section{Detailed Analysis of Benchmark Tasks \& Outliers}
\label{app:task-analysis}

Outlier solutions described below are found to be valid unless otherwise noted. We organize this analysis by breaking tasks and related outlier observations into the following categories:

\begin{itemize}
    \item Tasks which require refinement/removal based on our outlier analysis
    \item Tasks that we keep unchanged, but that lead to outlier results in certain cases (most of which we found to be valid upon manual inspection)
    \item Optimization techniques which can lead to timing challenges
\end{itemize}

\subsection{Task Refinement/Removal}

\subsubsection{LSTM and GRU Removal}

We opted to remove LSTM and GRU tasks entirely from Level 3-pike, as our LLM-generated solutions on these benchmarks frequently used CUDA Graphs to report $\sim$1000× speedups. While CUDA Graph solutions are not inherently invalid when using correct synchronization techniques, they do reveal glaring issues with these particular tasks entirely. 

We found that the PyTorch implementations of LSTM and GRU are poorly optimized for the chosen model size in KernelBench, launching each gate and each timestep as independent CUDA kernels for each layer of these operators. As a result, 1,000s of CUDA kernels are launched according to \verb|ncu| profiling tool, leading to a straw man baseline where kernel launch overhead dominates. Work could be done to extend the models or tune model hyperparameters, but we instead opted to remove these non-essential tasks.

\subsubsection{MLP Task Noise}

We found the Multilayer Perceptron (MLP) tasks in Level 3-pike (tasks 1 and 2) to be susceptible to noise during our analysis, given their small input and layer sizes. We find Triton's \verb|do_bench| utility to be much more stable for timing these smaller benchmarks compared to PyTorch's Timer utility, likely due to Triton's additional cache flushing and synchronization safety measures. Increasing batch size can also help to negate this issue. Though we optimized for batch size 1 in our evaluation, we recommend increasing batch size to 1,000 for future use of this benchmark.

\subsection{Level 3-pike Outliers}

The speedup breakdown for individual tasks on Level 3-pike can be found in Table~\ref{tab:breakdown-level3-pike}, including valid outlier speedups which are described here. We refer to the specific PIKE variants in this table as the PIKE implementations below.

\subsubsection{VisionAttention}

The PIKE-B VisionAttention solution achieves 28.67× on this benchmark through a combination of lowering precision to FP16, as well as replacing the built-in PyTorch \verb|nn.MultiheadAttention| with a custom Triton flash attention implementation. Numerous solutions find this combination of optimizations, but the 28.67× solution goes beyond that to include custom Triton kernels for fused input and output projection.

\subsubsection{Mamba2ReturnFinalState}

The PIKE implementations achieve speedups of 12.26 (PIKE-B) and 15.01 (PIKE-O ({\verb|mut,npar,1isl,EO,SL|})) by fusing all model operations into single, monolithic CUDA/Triton kernels, including calculating chunk sums, performing the inter-chunk scan, performing the intra-chunk scan, and accumulating final state. PIKE-B's fused kernel includes shared memory optimizations, vectorized accesses, and parallel scans implemented via warp-shuffle intrinsics, among other expert-level optimizations. 

\subsubsection{MinGPT}

The MinGPT examples benefit from lowering precision to FP16. These examples include a hand-written, unoptimized PyTorch implementation of causal self-attention, without any use of built-in PyTorch attention operations. As a result, the ideal optimization strategy is to identify how this unoptimized attention implementation can be replaced with a built-in PyTorch flash attention operator, or possibly a custom Triton-based attention implementation. METR successfully performs this combination of optimizations on both MinGPT tasks, achieving speedup of $\sim$12 on each. PIKE-O ({\verb|mut,npar,1isl,EO,SL|}) performs these optimizations as well in both cases, achieving similar performance to METR on both tasks. PIKE-B finds this optimal combination on 1 of the 2 MinGPT tasks.

\subsubsection{DenseNet121TransitionLayer}

The best PIKE solutions for this task find a valid transformation that reorders \verb|Conv2d| $\to$ \verb|AvgPool2d| operations to \verb|AvgPool2d| $\to$ \verb|Conv2d|. The convolution has kernel size $1 \times 1$, and the pool has kernel size $2 \times 2$, allowing the convolution operation to commute and reducing the number of elements it gets applied to. Combining this with precision reduction to FP16 and using custom fused kernels can yield speedup of $\sim$13.

\subsubsection{SqueezeNetFireModule}

PIKE-O ({\verb|mut,npar,1isl,EO,SL|}) achieves speedup of 8.56 on this task via a fused expand layer Triton kernel. The ReLU is also fused with the final convolutions in a separate Triton kernel, avoiding intermediate writes. This solution uses Triton's built-in autotuning as well.

\subsection{Level 5 Outliers}

The speedup breakdown for individual tasks on Level 5 can be found in Table~\ref{tab:breakdown-level5}, including valid outlier speedups which are described here. We refer to the specific PIKE variants in this table as the PIKE implementations below.

\subsubsection{S4}

On S4, the original model and the PIKE implementations take an FFT-based approach to the problem. METR abandons the FFT-based approach entirely. METR’s solution implements a direct, time-domain convolution inside a single monolithic kernel, gaining a substantial performance advantage over PIKE implementations to achieve speedup of 9.89.

\subsubsection{StableDiffusion3}

On StableDiffusion3, METR uses a fused Triton attention kernel which includes QKV projection, and also lowers precision to FP16, achieving speedup of 5.75. Our PIKE implementations stay in FP32 and use the built-in PyTorch attention operator, achieving lower speedups than METR.

\subsubsection{Mamba2}

Similar to the Level 3-pike Mamba2 components, the PIKE-B Mamba2 model for Level 5 is optimized by fusing all model scan operations into a single, monolithic CUDA kernel, achieving speedup of 10.81 in this case.

\subsubsection{DeepSeek3MOESmallBatch}

Both of our PIKE implementations optimize DeepSeek3MOESmallBatch (DeepSeek3MOEs) by eliminating the MOE expert computation for-loop, instead performing the expert computations all in parallel. The PIKE implementations pack expert weights into contiguous vectors to enable batch computations like batched/grouped matrix multiplication, executing one larger kernel for the matrix multiplication instead of many smaller ones. This results in speedup of $\sim$10 for the PIKE implementations.

\subsubsection{HunyuanVideo}

PIKE-B performs well on HunyuanVideoVAEDecoder (HunyuanDec) via a custom flash attention Triton kernel, with on-the-fly causal masking, similar to PIKE optimizations in Level 3-pike attention-based tasks, achieving speedup of 8.72 on this task.

Both PIKE implementations perform well on HunyuanVideoVAEEncoder (HunyuanEnc) by recognizing that the time embeddings can be removed entirely for the encoding path of this task, allowing unneeded branches to be eliminated. PIKE-O ({\verb|mut,npar,1isl,EO,SL|}) implements on-the-fly causal masking here, outperforming PIKE-B to achieve speedup of 9.74.

\subsubsection{Invalid Solutions}

We found an invalid METR solution for task DeepSeek3MLA, implementing rotary embedding as a no-op. We set the METR speedup on this task to 1, as no other METR solutions are available to fall back to.

\subsection{Challenging CUDA Techniques}

We found that for evaluation, CUDA Stream timing is only valid when \verb|torch.cuda.synchronize| is placed after each timing iteration, within the timed region. The same applies for CUDA Graphs. In our PIKE-O (no EFA) results for Level 3-pike, we found one result using \verb|torch.jit.trace| in combination with a custom CUDA kernel. We removed this result due to invalid timing measurements, likely as a result of internal synchronization issues.

\section{Artifact Appendix}

\subsection{Abstract}

This artifact describes how to reproduce the key results presented in this paper. Specifically, the steps of running an evaluation worker, running an LLM-based search strategy, and plotting results are documented. Details are provided on how to adjust the search strategy to use a different budget, select the desired LLM, and run on specific tasks. A host system with one H100 GPU, a CPU with 20+ threads, at least 64 GB RAM, and at least 200 GB storage are required for full, efficient results reproduction. Docker with the NVIDIA Container Toolkit should be available on the machine to run the evaluation worker container.

\subsection{Artifact check-list (meta-information)}

{\small
\begin{itemize}
  \item {\bf Program: KernelBench (public) with Python}
  \item {\bf Compilation: CUDA/Triton (in container)}
  \item {\bf Run-time environment: Linux 4.18+, Docker with the NVIDIA Container Toolkit (root needed to install), uv and Python 3.12}
  \item {\bf Hardware: one H100, CPU with 20+ threads, 64+ GB RAM}
  \item {\bf Execution: 24 hours for full run}
  \item {\bf Metrics: execution time}
  \item {\bf Output: Geomean speedup bar graph, cost vs. speedup graph}
  \item {\bf Experiments: see GitHub README}
  \item {\bf How much disk space required (approximately)?: 100 GB to reproduce subset of results (Docker container is 35 GB), 200 GB to reproduce fully}
  \item {\bf How much time is needed to prepare workflow (approximately)?: 30 minutes}
  \item {\bf How much time is needed to complete experiments (approximately)?: 2-4 hours single task, 24 hours full run (one search strategy, all tasks in a given KernelBench level), multiple days for many runs}
  \item {\bf Publicly available?: Yes}
  \item {\bf Code licenses (if publicly available)?: MIT (PIKE), Apache License 2.0 (PIKE OpenEvolve)}
  \item {\bf Data licenses (if publicly available)?: CC-BY-NC-4.0}
  \item {\bf Archived (provide DOI)?: \url{https://doi.org/10.5281/zenodo.18974037}}
\end{itemize}

\subsection{Description}

\subsubsection{How delivered}

Full artifact: \url{https://zenodo.org/records/18974037}

Main repository: \url{https://github.com/pike-project/pike}

The main repository has utility scripts to fetch the following: PIKE OpenEvolve (\url{https://github.com/pike-project/pike-openevolve}), data produced by our runs (\url{https://huggingface.co/datasets/knagaitsev/pike-data-compressed}), evaluation worker container image: (ghcr.io/knagaitsev/kernel-bench-deps)

\subsubsection{Hardware dependencies}

One NVIDIA H100 GPU, CPU with 20+ threads and 64+ GB RAM, to ensure the evaluation worker can compile many GPU kernels in parallel

\subsubsection{Software dependencies}

Docker with the NVIDIA Container Toolkit is needed for the evaluation worker container. uv is needed to install host dependencies and Python 3.12 for driving the search process

\subsubsection{Data sets}

Data for the runs included in our paper can be fetched from Hugging Face, including prompts, LLM responses, and runtimes. It is not necessary to fetch this data if one intends to reproduce the results entirely on their own.

\subsection{Installation}

Full instructions are provided in the main repository README. The outline is as follows:

\begin{enumerate}
    \item Have Docker with the NVIDIA Container Toolkit (requires root if not already installed)
    \item Install uv
    \item Create Python 3.12 virtual environment using uv
    \item Install repository requirements in virtual environment
\end{enumerate}

\subsection{Experiment workflow}

The outline from the main repository README is as follows:

\begin{enumerate}
    \item Start the evaluation worker Docker container using provided Python script (container image will be fetched from GitHub Container Registry)
    \item Run Python script on host to evaluate baselines by sending them to the evaluation worker (PyTorch eager, torch.compile, TensorRT, METR)
    \item Run Python script to start the search process on host, which submits code to the evaluation worker. Select desired search strategy, level/tasks, LLM, and query budget.
    \item Run figure generation Python script to plot results.
\end{enumerate}

\subsection{Evaluation and expected result}

The final figure generation script will produce geomean speedup plots, computing the geomean only for the tasks that were included in a given run. For a full run with a standard configuration (e.g. PIKE-B), we observed geomean speedup variation between runs to not exceed $\sim$10\% using the same LLM. Geomean speedup within this tolerance relative to the geomean provided in this paper is expected when using the specified hardware, same query budget (300 LLM queries), same strategy (e.g. PIKE-B), and same LLM.

\subsection{Experiment customization}

Details are provided in the README on how to adjust the search strategy to use a different budget, select the desired LLM, and run on specific tasks.

\subsection{Methodology}

Submission, reviewing and badging methodology:

\begin{itemize}
  \item \url{http://cTuning.org/ae/submission-20190109.html}
  \item \url{http://cTuning.org/ae/reviewing-20190109.html}
  \item \url{https://www.acm.org/publications/policies/artifact-review-badging}
\end{itemize}

\end{document}